\documentclass[aps,prd,onecolumn,groupedaddress,showpacs,nofootinbib,amssymb]{revtex4}
\usepackage{graphicx,bm}
\usepackage[english]{babel}
\usepackage{amsmath}
\usepackage{amssymb}
\usepackage{amsfonts}

\begin{document}

\def\cL{{\cal L}}
\def\be{\begin{equation}}
\def\ee{\end{equation}}
\def\bea{\begin{eqnarray}}
\def\eea{\end{eqnarray}}
\def\beq{\begin{eqnarray}}
\def\eeq{\end{eqnarray}}
\def\tr{{\rm tr}\, }
\def\nn{\nonumber \\}
\def\e{{\rm e}}


\title{
A proposal for covariant renormalizable field theory of gravity}

\author{Shin'ichi Nojiri$^{1,2}$ and Sergei D. Odintsov$^3$\footnote{
Also at Tomsk State Pedagogical University}}
\affiliation{
$^1$ Department of Physics, Nagoya University, Nagoya 464-8602, Japan \\
$^2$ Kobayashi-Maskawa Institute for the Origin of Particles and the Universe,\footnote{
See, http://www.kmi.nagoya-u.ac.jp/index-e.html .
}
Nagoya University, Nagoya 464-8602, Japan \\
$^3$Instituci\`{o} Catalana de Recerca i Estudis Avan\c{c}ats (ICREA)
and Institut de Ciencies de l'Espai (IEEC-CSIC),
Campus UAB, Facultat de Ciencies, Torre C5-Par-2a pl, E-08193 Bellaterra
(Barcelona), Spain
}

\begin{abstract}

The class of covariant gravity theories which have nice ultraviolet behavior
and seem to be (super)-renormalizable is proposed. The apparent breaking of
Lorentz invariance occurs due to the coupling with the effective fluid
which is induced by Lagrange multiplier constrained scalar field.
Spatially-flat FRW cosmology for such covariant field gravity may have
accelerating solutions. Renormalizable versions
of more complicated modified gravity
which depends on Riemann and Ricci tensor squared may be constructed in the
same way.

\end{abstract}


\maketitle


\noindent
{\bf Introduction.}
The main problem related with the quantization of the gravity
is that if we consider the perturbations from the flat background,
which has a Lorentz invariance, by using General Relativity, there appear
the non-renormalizable divergences  from the ultraviolet region in momentum
space.
Higher-derivative gravity may be renormalizable (see book \cite{ilb}) but the
well-known
unitarity problem cannot be solved there.
The idea proposed in ref.\cite{Horava:2009uw} (for
cosmological applications, see \cite{cosmology}) is to modify the ultraviolet
behavior
of the graviton propagator in Lorentz non-invariant way as
$1/\left|\bm{k}\right|^{2z}$,
where $\bm{k}$ is the spatial momenta and $z$ could be 2, 3 or larger integers.
  They are defined  by the scaling properties of space-time
coordinates $\left(\bm{x},t\right)$ as follows,
$\bm{x}\to b\bm{x}$, $t\to b^z t$.
When $z=3$, the theory seems to be UV renormalizable.
Then in order to realize the Lorentz non-invariance, one introduces the
terms breaking the Lorentz invariance explicitly (or more precisely, breaking
full
diffeomorphism invariance) by treating the temporal coordinate and the spatial
coordinates in a different  way.

Such model has invariance under time
reparametrization and time dependent spatial diffeomorphisms:
$\delta x^i=\zeta^i(t,\bm{x})$, $\delta t=f(t)$.
Here $\zeta^i(t,\bm{x})$ and $f(t)$ are arbitrary functions.

In ref.\cite{Nojiri:2009th}, Ho\v{r}ava-like gravity model with full
diffeomorphism invariance
has been proposed.
When we consider the perturbations from the flat background, which has Lorentz
invariance,
the Lorentz invariance of the propagator is dynamically broken by the
non-standard
coupling with a perfect fluid. The obtained propagator behaves as
$1/{\bm{k}}^{2z}$ with
$z=2,3,\cdots$ in the ultraviolet region and the model could be perturbatively
power counting
(super-)renormalizable if $z\geq 3$.
The price for such covariant renormalizability is the presence of unknown
(string-inspired?) fluid.
This fluid could not correspond to the usual fluid like, radiation, baryons,
dust, etc.
The model can be consistently constructed when the equation of state (EoS)
parameter $w\neq -1\, ,1/3$.
For usual particles in the high energy region, the corresponding fluid is
relativistic radiation
for which $w\to 1/3$. We need the non-relativistic fluid even in the high
energy region.

Recently dust fluid with $w=0$ has been constructed for the scalar theory by
introducing the Lagrange multiplier field, which gives a constraint
on the first scalar field \cite{Lim:2010yk}.
In this letter, we construct a fluid with arbitrary constant $w$ from the
scalar field which satisfies a
constraint. Due to the constraint, the scalar field is not dynamical and even
in the high energy
region, one can obtain a non-relativistic fluid. By the coupling with the
fluid,
one can get the full diffeomorphism invariant Lagrangian (actually, class of
such gravitational Lagrangians) given completely in terms of fields variables.
It is demonstrated that such theory has the good properties of Lorentz
non-invariant gravity
(its conjectured renormalizability) being on the same time the covariant one.
Moreover, in simplest case its spatially-flat FRW cosmology may have accelerating solutions.


\

\noindent
{\bf Review of covariant renormalizable gravity.}
Let us briefly review the covariant renormalizable gravity of
ref.\cite{Nojiri:2009th}.
The starting action is
\be
\label{Hrv1}
S = \int d^4 x \sqrt{-g} \left\{ \frac{R}{2\kappa^2} - \alpha \left( T^{\mu\nu}
R_{\mu\nu}
+ \beta T R \right)^2 \right\}\ .
\ee
Here $T_{\mu\nu}$ is  energy-momentum tensor of the exotic fluid.
The action (\ref{Hrv1}) is fully diffeomorphism invariant.
We consider the perturbation from the flat background $g_{\mu\nu} =
\eta_{\mu\nu} + h_{\mu\nu}$.
We now choose the following gauge conditions:
$h_{tt} = h_{ti} = h_{it} = 0$.
For the perfect fluid, the energy-momentum tensor in the flat background has
the following form:
\be
\label{Hrv5}
T_{tt} = \rho \ ,\quad T_{ij} = p \delta_{ij} = w \rho \delta_{ij}\ .
\ee
Here $w$ is the equation of state (EoS) parameter.
Then one finds
\bea
\label{Hrv6}
&& T^{\mu\nu} R_{\mu\nu} + \beta T R \nn
&& = \rho \left[ \left\{ - \frac{1}{2} + \frac{w}{2} + \left( - 1 + 3w \right)
\beta \right\}
\partial_t^2 \left(\delta^{ij} h_{ij} \right)
+ \left( w - \beta + 3w \beta \right) \partial^i \partial^j h_{ij}
+ \left( - w + \beta - 3w \beta \right)
\partial_k \partial^k \left(\delta^{ij} h_{ij} \right) \right]
\eea
If we choose
\be
\label{Hrv7}
\beta = - \frac{w-1}{2\left(3w - 1\right)}\ ,
\ee
the second term in the action (\ref{Hrv1}) becomes
\be
\label{Hrv8}
\alpha \left( T^{\mu\nu} R_{\mu\nu} + \beta T R \right)^2
= \alpha \rho^2 \left(\frac{w}{2} + \frac{1}{2} \right)^2 \left\{  \partial^i
\partial^j h_{ij}
  - \partial_k \partial^k \left(\delta^{ij} h_{ij} \right) \right\}^2\ ,
\ee
which does not contain the derivative with respect to $t$ and breaks the
Lorentz invariance.
We now assume $\rho$ is almost constant.
Then in the ultraviolet region, where $\bm{k}$ is large, the second term in the
action (\ref{Hrv1})
gives the propagator behaving as $1/\left| \bm{k} \right|^4$,
which renders the ultraviolet behavior (compared with Eq.(1.4) in
\cite{Horava:2009uw}).
Note that the form (\ref{Hrv7}) indicates that the longitudinal mode does
not propagate but only the transverse mode propagates.

There are two special cases in the choice of $w$: when $w=-1$, which
corresponds to the cosmological constant, one gets
$T^{\mu\nu} R_{\mu\nu} + \beta T R=0$ and
therefore we do not obtain $1/\bm{k}^4$ behavior. When $w=1/3$, which
corresponds
to the radiation or conformal matter, $\beta$ diverges and therefore there
is no solution.

The apparent breakdown of the Lorentz symmetry in (\ref{Hrv8}) occurs
due to the coupling with the perfect fluid.
The action (\ref{Hrv1}) is invariant under the diffeomorphisms
in four dimensions and the energy-momentum tensor $T_{\mu\nu}$ of the
non-standard fluid
in the action should transform as a tensor under the diffeomorphisms.
The existence of the fluid, however, effectively breaks the Lorentz symmetry,
which is the equivalence between the different inertial frames of reference.
Note that the expression (\ref{Hrv5}) is correct in the reference frame where
the fluid does not flow,
or the velocity of the fluid vanishes. In other reference frames, there
appear non-vanishing $T_{it}=T_{ti}$ components and there could appear the
derivative
with respect to time, in general.

In the arguments after (\ref{Hrv1}),  the flat background is considered
but the arguments could be generalized for the curved background: in
the
curved
spacetime, due to the principle of the general relativity, we can always choose
the local
Lorentz frame. The local Lorentz frame has (local) Lorentz symmetry. Even in
the local Lorentz
frame, the perfect fluid might flow and $T_{it}=T_{ti}$ components might not
vanish.
By boosting the frame, which is the (local)
Lorentz transformation, we have a special local Lorentz frame, where the fluid
does not flow.
In the Lorentz frame, one can use the above arguments and find there is no
breakdown
of the unitarity. Conversely, in a general coordinate frame, $T^{\mu\nu}
R_{\mu\nu} + \beta T R$
can have a derivative with respect to time.

The action  (\ref{Hrv1}) gives $z=2$ theory.
In order that the theory could be ultra-violet power counting renormalizable in
$3+1$
dimensions,  $z=3$ theory is necessary. In order to obtain such a theory, we
note that,
for any scalar quantity $\Phi$, if we choose
\be
\label{Hrv13}
\gamma = \frac{1}{3 w - 1}\ ,
\ee
one obtains
\be
\label{Hrv14}
T^{\mu\nu}\nabla_\mu \nabla_\nu \Phi + \gamma T \nabla^\rho \nabla_\rho \Phi
= \rho \left( w + 1 \right) \partial_k \partial^k \Phi \ ,
\ee
which does not contain the derivative with respect to time coordinate $t$.
This is true even if the coordinate frame is not local Lorentz frame.
The derivative with respect to time coordinate $t$ is not contained in any
coordinate
frame, where the perfect fluid does not flow.
Then if we consider
\be
\label{Hrv15}
S = \int d^4 x \sqrt{-g} \left\{ \frac{R}{2\kappa^2} - \alpha \left( T^{\mu\nu}
R_{\mu\nu}
+ \beta T R \right)
\left(T^{\mu\nu}\nabla_\mu \nabla_\nu + \gamma T \nabla^\rho \nabla_\rho\right)
\left( T^{\mu\nu} R_{\mu\nu} + \beta T R \right)
\right\}\ ,
\ee
with $\beta = - \frac{w-1}{2\left(3w - 1\right)}$ and
$\gamma = \frac{1}{3 w - 1}$,
we obtain $z=3$ theory, which seems to be renormalizable.
In general, for the case
\be
\label{Hrv18}
S = \int d^4 x \sqrt{-g} \left[ \frac{R}{2\kappa^2} - \alpha \left\{
\left(T^{\mu\nu}\nabla_\mu \nabla_\nu + \gamma T \nabla^\rho
\nabla_\rho\right)^n
\left( T^{\mu\nu} R_{\mu\nu} + \beta T R \right) \right\}^2 \right]\ ,
\ee
with a constant $n$, we obtain $z = 2 n + 2$ theory which is
super-renormalizable for $n\geq 1$.
Usually $n$ should be an integer but in general, we may consider
pseudo-local differential
operator $\left(T^{\mu\nu}\nabla_\mu \nabla_\nu + \gamma T \nabla^\rho
\nabla_\rho\right)^n$
with non-integer $n$ (e.g. $n=1/2,3/2$ etc.).
We should also note that there are special cases, that is,
$w=-1$ and $w=1/3$.

The second terms in the actions (\ref{Hrv1}), (\ref{Hrv15}), (\ref{Hrv18}), and
(\ref{Hrv18}), which effectively break the Lorentz symmetry, are relevant only
in the high
energy/UV region since they contain higher derivative terms. In the IR
region, these terms do not dominate and the usual Einstein gravity follows as a
limit.

In \cite{Kluson:2010xx,  Chaichian:2010yi}, $F(R)$-gravity \cite{review} version
of the Ho\v{r}ava-Lifshitz-like gravity has been proposed. The renormaliability
of this
class of $F(R)$-gravity has been established in \cite{Carloni:2010nx}.
In \cite{Carloni:2010nx}, the covariant version of the Ho\v{r}ava-Lifshitz-like
$F(R)$-gravity has been also presented. Its action is given by
\bea
\label{KLFRg24}
S_{F(\tilde R_\mathrm{cov} )} &=& \frac{1}{2\kappa^2}\int d^4 x
\sqrt{- g} F(R_\mathrm{cov})\, , \nn
R_\mathrm{cov} &=& \left\{
\begin{array}{ll}
R - 2 \alpha \kappa^2 \left( T^{\mu\nu} R_{\mu\nu}
+ \beta T R \right)^2\, , & z=2 \, ,\\
R - 2\kappa^2 \alpha \left( T^{\mu\nu} R_{\mu\nu}
+ \beta T R \right)
\left(T^{\mu\nu}\nabla_\mu \nabla_\nu + \gamma T \nabla^\rho
\nabla_\rho\right)
\left( T^{\mu\nu} R_{\mu\nu} + \beta T R \right) & z=3 \, ,\\
R - 2\kappa^2 \alpha \left\{
\left(T^{\mu\nu}\nabla_\mu \nabla_\nu + \gamma T \nabla^\rho
\nabla_\rho\right)^n
\left( T^{\mu\nu} R_{\mu\nu} + \beta T R \right)\right\}^2 & z = 2 n + 2 \, .\\
\end{array}
\right.
\eea
Although it has not been shown, the theories
(\ref{KLFRg24}) are expected to
be renormalizable when $z\geq 3$, which may be demonstrated using the
same arguments as in ref.~\cite{Nojiri:2009th} or in this section.


\

\noindent
{\bf Covariant renormalizable field theory of gravity with Lagrange
multiplier.}
In this section, the covariant 
gravity coupled with the (Lagrange multiplier induced)fluid  is
constructed.

We consider the following constrained action for the scalar field $\phi$
\be
\label{LagHL1}
S_\phi = \int d^4 x \sqrt{-g} \left\{ - \lambda \left( \frac{1}{2}
\partial_\mu \phi \partial^\mu \phi + U(\phi) \right) \right\}  \, .
\ee
Here $\lambda$ is the Lagrange multiplier field, which gives a constraint
\be
\label{LagHL2}
\frac{1}{2} \partial_\mu \phi \partial^\mu \phi
+ U(\phi) = 0\, ,
\ee
that is, the vector $(\partial_\mu \phi)$ is time-like.
At least locally, one can choose the direction of time to be parallel to
$(\partial_\mu \phi)$.
Then Eq. (\ref{LagHL2}) has the following form:
\be
\label{LagHL3}
\frac{1}{2} \left(\frac{d\phi}{dt}\right)^2 = U(\phi)\, .
\ee
The equation given by the variation of $\phi$ will be discussed later.

We now {\it define} a tensor $T^\phi_{\mu\nu}$ corresponding to the energy
momentum tensor of the scalar field with a potential $V(\phi)$:
\be
\label{LagHL4}
T^\phi_{\mu\nu} = \partial_\mu \phi \partial_\nu \phi
  - g_{\mu\nu} \left( \frac{1}{2} \partial_\rho \phi \partial^\rho \phi
+ V(\phi) \right) \, .
\ee
The ``energy density'' $\rho_\phi$ and ``pressure'' $p_\phi$ become:
\be
\label{LagHL5}
p_\phi = \frac{1}{2} \left(\frac{d\phi}{dt}\right)^2 - V(\phi) = U(\phi)  -
V(\phi) \, ,\quad
\rho_\phi = \frac{1}{2} \left(\frac{d\phi}{dt}\right)^2 + V(\phi) = U(\phi)  +
V(\phi) \,  .
\ee
Here, the constraint (\ref{LagHL3}) is used.
We should note that $V(\phi)$ is not identical with $U(\phi)$: $V(\phi)\neq
U(\phi)$.
In case $V(\phi) = U(\phi)$, Eq.(\ref{LagHL5}) tells that $p_\phi =0$, which
corresponds to
dust with $w_\phi \equiv p_\phi/\rho_\phi = 0$. Note that quantization of
constrained theories
is quite non-trivial task (see reviews\cite{masud}).

For simplicity, we choose $V(\phi)$ and $U(\phi)$ to be constants:
\be
\label{LagHL6}
U(\phi) = U_0\, ,\quad V(\phi) = V_0\, .
\ee
Then if $U_0 = V_0$, the EoS parameter $w_\phi$ vanish. In general case,
one has $w_\phi = \frac{U_0 - V_0}{U_0 +V_0}$.
Let us now use $T^\phi_{\mu\nu}$ as a energy-momentum tensor in the previous
section.
Since (\ref{Hrv7}) shows
$\beta = - \frac{w-1}{2\left(3w - 1\right)}
= \frac{V_0}{2U_0 - 4V_0}$.
One can simplify
\be
\label{HrvHL9}
T^{\phi\, \mu\nu} R_{\mu\nu} + \beta T^\phi R
= \partial^\mu \phi \partial^\nu \phi R_{\mu\nu} + U_0 R\, .
\ee
Here, Eqs.(\ref{LagHL2}), (\ref{LagHL4}), and (\ref{LagHL6}) are used.
We also find $\gamma$ in (\ref{Hrv13}) has the following form:
$\gamma = \frac{U_0 - V_0}{2U_0 - 4V_0}$,
which gives, by using (\ref{Hrv14}),
\be
\label{HrvHL10}
T^{\phi\,\mu\nu}\nabla_\mu \nabla_\nu \Phi + \gamma T^\phi \nabla^\rho
\nabla_\rho \Phi
= \partial^\mu \phi \partial^\nu \phi \nabla_\mu \phi \nabla_\nu \Phi
+ 2 U_0 \nabla^\rho \nabla_\rho \Phi\, .
\ee
Eq. (\ref{HrvHL9}) enables to write down $z=2$ total action corresponding to
(\ref{Hrv1}) as
\be
\label{HrvHL11}
S = \int d^4 x \sqrt{-g} \left\{ \frac{R}{2\kappa^2}
 - \alpha \left( \partial^\mu \phi \partial^\nu \phi R_{\mu\nu} + U_0 R
\right)^2
 - \lambda \left( \frac{1}{2} \partial_\mu \phi \partial^\mu \phi
+ U_0 \right)
\right\} \, .
\ee
On the other hand, $z=3$ total action corresponding to (\ref{Hrv15}) has
the following form:
\bea
\label{HrvHL12}
S &=& \int d^4 x \sqrt{-g} \left\{ \frac{R}{2\kappa^2} - \alpha \left(
\partial^\mu \phi \partial^\nu \phi R_{\mu\nu} + U_0 R \right)
\left(\partial^\mu \phi \partial^\nu \phi \nabla_\mu \nabla_\nu
+ 2 U_0 \nabla^\rho \nabla_\rho \right)
\left( \partial^\mu \phi \partial^\nu \phi R_{\mu\nu} + U_0 R \right) \right. \nn
&& \left. - \lambda \left( \frac{1}{2} \partial_\mu \phi \partial^\mu \phi
+ U_0 \right) \right\}\, ,
\eea
and $z = 2 n + 2$ action is given by
\bea
\label{HrvHL13}
S &=& \int d^4 x \sqrt{-g} \left[ \frac{R}{2\kappa^2} - \alpha \left\{
\left(\partial^\mu \phi \partial^\nu \phi \nabla_\mu \nabla_\nu
+ 2 U_0 \nabla^\rho \nabla_\rho \right)^n
\left( \partial^\mu \phi \partial^\nu \phi R_{\mu\nu} + U_0 R \right) \right\}^2 \right. \nn
&& \left. - \lambda \left( \frac{1}{2} \partial_\mu \phi \partial^\mu \phi
+ U_0 \right) \right]\, ,
\eea
Note that the actions (\ref{HrvHL11}), (\ref{HrvHL12}), and (\ref{HrvHL13})
are totally diffeomorphism invariant and only given in terms of the local
fields. We also note that the actions (\ref{HrvHL11}), (\ref{HrvHL12}), and (\ref{HrvHL11})
do not depend on $V_0$.

By the variation over $\phi$,
for example, for $z=2$ case in (\ref{HrvHL11}),one finds
\be
\label{CRG_phi1}
0 = 4 \alpha \partial^\mu \left\{ \partial^\nu \phi R_{\mu\nu}
\left( \partial^\rho \phi \partial^\sigma \phi R_{\rho\sigma} + U_0 R \right)
\right\} +  \partial^\mu \left( \lambda \partial_\mu \phi \right)  \, .
\ee
For $z=3$ case (\ref{HrvHL12}) or $z=2n+2$ case (\ref{HrvHL13}),
 rather complicated equations follow.


For the perturbation from the flat background
$g_{\mu\nu} = \eta_{\mu\nu} + h_{\mu\nu}$, we find
\bea
\label{HrvHL14}
\partial^\mu \phi \partial^\nu \phi R_{\mu\nu} + U_0 R &=&
U_0 \left\{  \partial^i \partial^j h_{ij}
 - \partial_k \partial^k \left(\delta^{ij} h_{ij} \right) \right\}\, , \nn
\partial^\mu \phi \partial^\nu \phi \nabla_\mu \nabla_\nu
+ 2 U_0 \nabla^\rho \nabla_\rho &=&
2U_0 \partial_k \partial^k \, .
\eea
Then in the ultraviolet region, where $\bm{k}$ is large,
the propagator behaves as $1/\left| \bm{k} \right|^4$ for $z=2$ case in
(\ref{HrvHL11}) and therefore the ultraviolet behavior is rendered.
In $z=3$ case in (\ref{HrvHL12}),
the propagator behaves as $1/\left| \bm{k} \right|^6$ and therefore
the model becomes renormalizable.
In $z=2n +2$ case in (\ref{HrvHL13}), when $n\geq 1$, the model becomes
super-renormalizable.

The $F(R)$-gravity corresponding to (\ref{KLFRg24}) is given by
\bea
\label{HrvHL15}
S_{F(\tilde R_\mathrm{cov} )} &=& \frac{1}{2\kappa^2}\int d^4 x
\sqrt{- g} \left\{ F(R_\mathrm{cov})
  - \lambda \left( \frac{1}{2} \partial_\mu \phi \partial^\mu \phi
+ U_0 \right) \right\}\, ,\nn
R_\mathrm{cov} &=& \left\{
\begin{array}{ll}
R - 2 \alpha \kappa^2 \left(
\partial^\mu \phi \partial^\nu \phi R_{\mu\nu} + U_0 R \right)^2\, , & z=2 \, ,\\
R - 2\kappa^2 \alpha \left(
\partial^\mu \phi \partial^\nu \phi R_{\mu\nu} + U_0 R \right)
\left( \partial^\mu \phi \partial^\nu \phi \nabla_\mu \nabla_\nu
+ 2 U_0 \nabla^\rho \nabla_\rho \right)
\left( \partial^\mu \phi \partial^\nu \phi R_{\mu\nu} + U_0 R
\right) & z=3 \, ,\\
R - 2\kappa^2 \alpha \left\{
\left( \partial^\mu \phi \partial^\nu \phi \nabla_\mu \nabla_\nu
+ 2 U_0 \nabla^\rho \nabla_\rho \right)^n
\left( \partial^\mu \phi \partial^\nu \phi R_{\mu\nu} + U_0 R
\right)\right\}^2 & z = 2 n + 2 \, .\\
\end{array}
\right.
\eea
The action (\ref{HrvHL15}) is also totally diffeomorphism invariant and only
given in terms of the local fields.


\

\noindent
{\bf Discussion: cosmological applications.}
Let us make several remarks about FRW cosmology in the presence of matter.
In order to obtain  the FRW equations, we assume the following
form of the metric:
$ds^2 = - \e^{2b(t)}dt^2 + a(t)^2 \sum_{i=1,2,3} \left(dx^i\right)^2$,
and  that the scalar field $\phi$ only depends on time.
Then
\be
\label{HrvCos1}
\partial^\mu \phi \partial^\nu \phi R_{\mu\nu} + U_0 R = 6H^2 U_0 \e^{-2b}\, ,\quad
\partial^\mu \phi \partial^\nu \phi \nabla_\mu \nabla_\nu
+ 2 U_0 \nabla^\rho \nabla_\rho = - 6 U_0 \e^{-2b} H \partial_t\, .
\ee
For simplicity, we consider the model with $z=2$  (\ref{HrvHL11}), then
the action (\ref{HrvHL11}) has the following form:
\be
\label{Hrv19d}
S = \int d^4 x a^3 \left[ \frac{\e^{-b}}{2\kappa^2} \left(6\dot H + 12 H^2 -
6\dot b H\right)
 - 36\alpha\ U_0^2 \e^{-3b} H^4
 - \lambda \left( - \frac{\e^{-b}}{2}{\dot \phi}^2 + \e^b U_0 \right) \right]\, .
\ee
The equation corresponding to the first FRW equation can be obtained by putting
$b=0$ after the variation over $b$ and it has the following form:
\be
\label{HrvHL16}
\frac{3}{\kappa^2} H^2 = - 108 \alpha U_0^2 H^4 + 2 \lambda U_0 + \rho_\mathrm{matter} \, .
\ee
Here we have used a constraint (\ref{LagHL3}), which is valid even in the FRW
universe and  the usual matter energy-density $\rho_\mathrm{matter}$ is included.
On the other hand, by considering the variation over $a$ and putting $b=0$, one
obtains the equation corresponding to the second FRW equation for (\ref{HrvHL12}):
\be
\label{HrvHL17}
 - \frac{1}{\kappa^2} \left( 2\dot H + 3 H^2 \right)
= 36 \alpha U_0^2 \left( 3H^4 + 4H^2 \dot H \right)
+ p_\mathrm{matter} \, .
\ee
Here we have used a constraint (\ref{LagHL3}), again
and $p_\mathrm{matter}$ is the usual matter pressure.
Note that Eq. (\ref{HrvHL17}) does not contain $\lambda$. By solving (\ref{HrvHL17}),
$H=H(t)$. After that, by substituting the solution $H(t)$, we can find the form of $\lambda$.

In the early universe with large curvature, the contribution from the Einstein term,
which corresponds to the right-hand sides in (\ref{HrvHL16}) and (\ref{HrvHL17}) is large, 
and the matter contribution could be neglected.
Then a solution of (\ref{HrvHL17}) is given by
\be
\label{sym41}
H=\frac{4}{3t}\, ,
\ee
which expresses the (power law) accelerating expansion of the universe corresponding
to the perfect fluid with $w=-1/2$.
Then Eq.(\ref{HrvHL16}) gives
\be
\label{sym42}
\lambda = \frac{32 \alpha U_0}{3t^4}\, .
\ee
This accelerated FRW cosmology may be proposed to describe (quintessencial) inflationary era.

One may confirm that the actions (\ref{HrvHL11}), (\ref{HrvHL12}), and (\ref{HrvHL13})
admit a  solution where $R=R_{\mu\nu}=0$, which corresponds to the flat,
Schwarzschild, or Kerr space-time, by investigating the Einstein equation:
\be
\label{sym4}
0 = \frac{1}{2\kappa^2} \left( R_{\mu\nu} - \frac{1}{2} g_{\mu\nu} R \right) + G^\mathrm{higher}_{\mu\nu}
 - \frac{\lambda}{2} \partial_\mu \phi \partial_\nu \phi + \frac{1}{2} g_{\mu\nu}
\left( \frac{1}{2} \partial_\rho \phi \partial^\rho \phi + U_0 \right)\, .
\ee
Here $G^\mathrm{higher}_{\mu\nu}$ comes from the higher derivative term (the second term)
in the action.
When  $R=R_{\mu\nu}=0$, then $G^\mathrm{higher}_{\mu\nu}=0$,
by using the constraint equation (\ref{LagHL2}), we find that Eq.(\ref{sym4}) reduces to
\be
\label{sym5}
0 = \lambda \partial_\mu \phi \partial_\nu \phi\, ,
\ee
whose solution is $\lambda=0$. Then the actions
(\ref{HrvHL11}), (\ref{HrvHL12}), and (\ref{HrvHL13}) admit the solution with
$R=R_{\mu\nu}=0$, which includes the Schwarzschild solution
\be
\label{sym28}
ds^2 = - \left( 1 - \frac{M}{r} \right) dt^2 + \left( 1 - \frac{M}{r} \right)^{-1} dr^2
+ r^2 d\Omega^2\, .
\ee
In the  Ho\v{r}ava gravity and the theories which we are considering, the dispersion
relation of the graviton is given by
\be
\label{sym29}
\omega = c_0 k^z\, ,
\ee
in the high energy region.
Here $c_0$ is a constant, $\omega$ is the angular frequency corresponding to the energy
and $k$ is the wave number corresponding to momentum.
Then the phase speed $v_\mathrm{p}$ and the group speed $v_\mathrm{g}$ are given by
\be
\label{sym30}
v_\mathrm{p} = \frac{\omega}{k} =  c_0 k^{z-1}\, ,\quad
v_\mathrm{g} = \frac{d\omega}{dk} =  c_0 z k^{z-1}\, ,
\ee
which becomes larger and larger when $k$ becomes larger and goes beyond the light speed.
This shows that even in (\ref{sym28}), the high energy graviton can escape from the horizon.
Note that the horizon is null surface and therefore in the usual Einstein gravity,
particle cannot escape from the horizon since the speed of the particle is always less than or equal
to the light speed. In our model, however, the speed of the graviton can exceed the light speed and
escape from the horizon.
This indicates that some properties of black holes in the theory under consideration are similar 
to the ones of Ho\v{r}ava gravity.


As a remark, instead of (\ref{HrvHL15}), one may investigate
the $F(R)$ type model where the action is given by
\be
\label{HrvHL18}
S_{R + F(\tilde R_\mathrm{cov} )} = \frac{1}{2\kappa^2}\int d^4 x
\sqrt{- g} \left\{ \frac{R}{2\kappa^2} + F(R_\mathrm{cov})
  - \lambda \left( \frac{1}{2} \partial_\mu \phi \partial^\mu \phi
+ U_0 \right) \right\}\, .
\ee
Here $R$ is the usual scalar curvature.
Note, however, the model (\ref{HrvHL18}) is not always renormalizable.
Such examples are $F(R_\mathrm{cov}) \propto R_\mathrm{cov}^n$ $(n\geq 3)$.
In order to investigate the renormalizability, let us consider the fluctuation
from the flat background.
Then  $R_\mathrm{cov}^n$ $(n\geq 3)$ term does not contain the second power of
the fluctuation
but only contains $n$-th or higher power terms like $h^m$ $(n\geq m)$. Then the
$R_\mathrm{cov}^n$ term
does not give any contribution to the propagator and therefore the ultraviolet
structure of
the divergence is never improved and there appear non-renormalizable
divergences.

Another remark is about the emergence of the standard Newton law.
In the original Ho\v{r}ava model \cite{Horava:2009uw},
the lapse function $N$ is restricted to only depend on the time coordinate,
which is called ``projectability condition''. This condition could be natural
since the original Ho\v{r}ava model has not full diffeomorphism invariance but
the invariance under time reparametrization and time dependent spatial diffeomorphisms.
As pointed out in \cite{Blas:2009qj}, by imposing the projectability condition, the Newton
law could not be reproduced. Even in the Ho\v{r}ava model, if the projectability condition
is not imposed, the Newton law could be realized.
The model proposed in this paper, however, has the full diffeomorphism invariance and
therefore we need not to impose the the projectability condition.
In the models (\ref{HrvHL11}), (\ref{HrvHL12}), and (\ref{HrvHL13}), the corrections to the
Einstein gravity come from the second and third terms.
The second term is the higher derivative term and therefore relevant only for the very short
distance, which possibly corresponds to the Planck length.
Then this term does not affect the Newton law which has not been checked for such
a short distance. The third term only gives a constraint and irrelevant for the Newton law.

In summary, we proposed class of covariant gravity theories which have nice
ultraviolet behavior and seem to
be (super)-renormalizable in the same sense as Ho\v{r}ava gravity which is
known to be Lorentz non-invariant.
These covariant theories are coupled with some fluid which is induced by the
corresponding Lagrange multiplier constrained scalar.
The accelerating spatially-flat FRW unified cosmology may be constructed for $F(R)$
versions of such theory, which opens the bridge between modified gravity cosmology
and renormalizability. Moreover, renormalizable versions of more complicated modified
gravity may be constructed in the same way.


\

\noindent
{\bf Acknowledgments}
This research has been supported in part by MEC (Spain) project FIS2006-02842
and AGAUR(Catalonia)
2009SGR-994 (SDO), by Global COE Program of Nagoya University (G07)
provided by the Ministry of Education, Culture, Sports, Science \& Technology (SN).

\end{document}